\documentclass[prl,aps,twocolumn,superscriptaddress,showpacs]{revtex4-1}

\usepackage{graphicx}% Include figure files
\usepackage{epstopdf}

\begin{document}

\title{Single to multi quasiparticle excitations in the itinerant helical magnet CeRhIn$_{5}$}

\author{C. Stock}
\affiliation{School of Physics and Astronomy, University of Edinburgh, Edinburgh EH9 3JZ, UK}
\author{J. A. Rodriguez-Rivera}
\affiliation{NIST Center for Neutron Research, National Institute of Standards and Technology, 100 Bureau Dr., Gaithersburg, MD 20899}
\affiliation{Department of Materials Science, University of Maryland, College Park, MD  20742}
\author{K. Schmalzl}
\affiliation{Julich Centre for Neutron Science, Forschungszentrum Julich GmbH, Outstation at Institut Laue-Langevin, Boite Postale 156, 38042 Grenoble Cedex 9, France}
\author{E. E. Rodriguez}
\affiliation{Department of Chemistry of Biochemistry, University of Maryland, College Park, MD, 20742, U.S.A.}
\author{A. Stunault}
\affiliation{Institute Laue-Langevin, B. P. 156, 6 rue Jules Horwitz, F-38042 Grenoble Cedex 9, France}
\author{C. Petrovic}
\affiliation{Department of Physics, Brookhaven National Laboratory, Upton, New York, 11973, USA}

\date{\today}

\begin{abstract}

CeRhIn$_{5}$ is an itinerant magnet where the Ce$^{3+}$ spins order in a simple helical phase.  We investigate the spin excitations and observe sharp spin-waves parameterized by a nearest neighbor exchange $J_{RKKY}$=0.88 $\pm$ 0.05 meV.  At higher energies, the spin fluctuations are heavily damped where single quasiparticle excitations are replaced by a momentum and energy broadened continuum constrained by kinematics of energy and momentum conservation.    The delicate energy balance between localized and itinerant characters results in the breakdown of the single quasiparticle picture in CeRhIn$_{5}$.

\end{abstract}


\maketitle

The noninteracting quasiparticle description of excitations is fundamental to condensed matter physics and the understanding of low energy fluctuations.  However, interacting quasiparticle states have recently been recognized as important for the understanding of anomalous phases. For example, composite states including resonating valence bond states~\cite{Anderson00:62}, Zhang-Rice singlets~\cite{Zhang88:37} or spinon-holons in the pseudogap,~\cite{Efetov13:9} have been suggested to be fundamental to superconductivity, frustrated magnetism, and even quantum criticality.~\cite{Han12:492,Vries09:103,Coleman01:13}  We use neutron scattering to measure the breakdown of the single quasiparticle description of the spin excitations in a helical itinerant heavy fermion magnet.
 
CeRhIn$_{5}$ is a heavy fermion metal, part of the Ce$T$In$_{5}$ ($T$=Rh, Ir, and Co) series displaying an interplay between localized antiferromagnetism and superconductivity.~\cite{Thompson01:226,Park09:11,Paglione05:94,Park08:105}  The presence of two-dimensional layers of Ce$^{3+}$ ions connects the physics of these systems with other unconventional superconductors as in the cuprates~\cite{Fujita12:81,Kastner98:70,Birgeneau06:75,Petrovic01:13,Hall01:64} or iron based pnictide/chalcogenide superconductors.~\cite{Stewart11:83,Paglione10:6,Johnston10:6}  CeRhIn$_{5}$ magnetically orders at T$_{N}$=3.8 K \cite{Bao00:62,Bao02:65,Raymond07:19} and enters an unconventional superconducting phase that can be accessed under hydrostatic pressures or temperatures below $\sim$ 75 mK.~\cite{Chen06:97,Paglione08:77,Park08:456,Park12:108,Ferreira08:101}  

CeRhIn$_{5}$ is isostructural with CeCoIn$_{5}$, which is superconducting at ambient pressures with a T$_{c}$=2.3 K.~\cite{Petrovic01:13}   The order parameter of the superconducting phase has a $d$-wave symmetry with nodes in the $ab$ plane.~\cite{Izawa01:87,Aoki04:16}  Magnetism and superconductivity are strongly coupled as evidenced by neutron scattering measurements reporting a doublet spin-resonance peak connected with superconductivity and indicating an order parameter that changes sign, consistent with $d$-wave symmetry.~\cite{Stock12:109,Stock08:100,Raymond12:109}  At high magnetic fields near H$_{c2}$, an unusual magnetic ``Q-phase" has been reported to exist in a narrow field region further confirming the interplay between superconductivity and the localized magnetism.~\cite{Kenzelmann08:321,Blackburn10:105}

\begin{figure}[t]
\includegraphics[width=7.5cm] {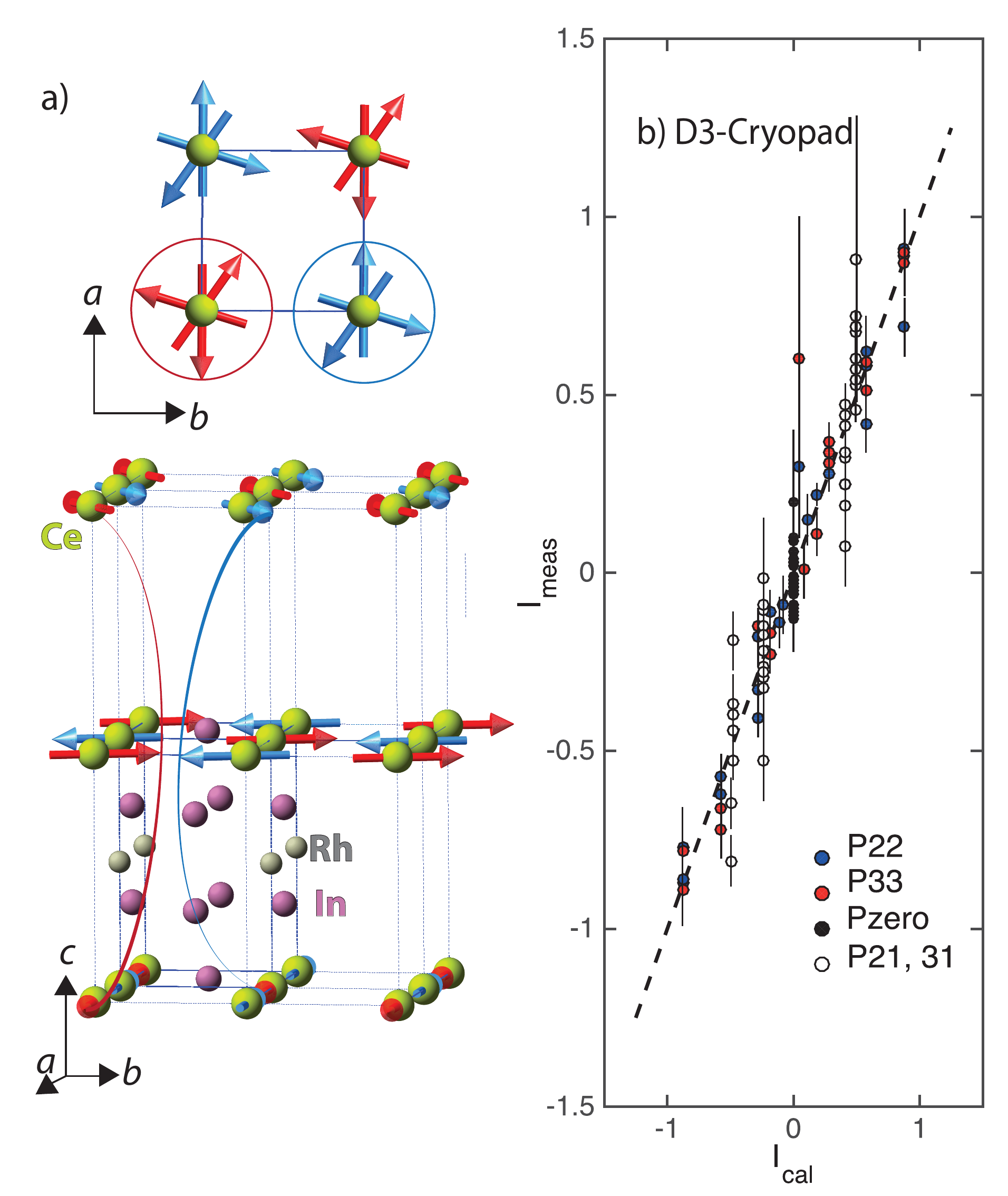}
\caption{\label{structure} The magnetic structure of CeRhIn$_{5}$ investigated using spherical polarimetry. $(a)$ An illustration of the magnetic structure.  $(b)$ is a plot of $P_{measured}$ vs $P_{calculated}$ based upon the isotropic helical model shown in $(a)$.}
\end{figure}

Neutron measurements were performed at NIST (Gaithersburg, USA) using MACS~\cite{Rodriguez08:19} and at the ILL (Grenoble, France) using the IN12 spectrometer and the D23 and D3 diffractometers. The $HHL$ aligned sample was prepared using self-flux method.~\cite{Petrovic01:13}  To correct for the large neutron absorption,~\cite{Sears92:3,Wuensch65:122} a finite element analysis has been done.  Further details are provided in the supplementary information.  

We first review the low temperature magnetic structure using spherical polarimetry.~\cite{Tasset99:267,Blume63:130,Brown93:442}  As found in the pioneering work by Bao \textit{et al.}~\cite{Bao00:62}, the magnetic structure (Fig. \ref{structure} (a)) is characterized by an incommensurate Bragg peak {\bf{Q}}=(0.5,0.5,0.297).  Fig. \ref{structure} (b) plots the results of our polarized diffraction experiment confirming this with measured polarization matrix elements (P$_{measured}$) against calculated (P$_{calculated}$) assuming a perfect $a-b$ helix with the moment defined by ${\bf{M}}={\bf{M}}_{a}+i{\bf{M}}_{b}$ (with $|{\bf{M_{a}}}|=|{\bf{M_{b}}}|$) and propagation vector along $c$.  Expressions for the matrix elements are given in the supplementary information.  Confirming the helical magnetism, a volume imbalance between the two chiral domains $\eta$=0.68 $\pm$ 0.05 was needed to account for off-diagonal matrix elements.  Unpolarized diffraction measures the ordered magnetic moment to be 0.34 $\pm$ 0.05 $\mu_{B}$ per cerium ion, consistent with expectations from crystal field theory.~\cite{Hutchings65:45}  The derived magnetic structure and symmetry analysis is also consistent with predictions from Landau theory for the phase transition as outlined in the supplemental information.~\cite{Dimmock63:130,Franzen90:2}

\begin{figure}[t]
\includegraphics[width=8.5cm] {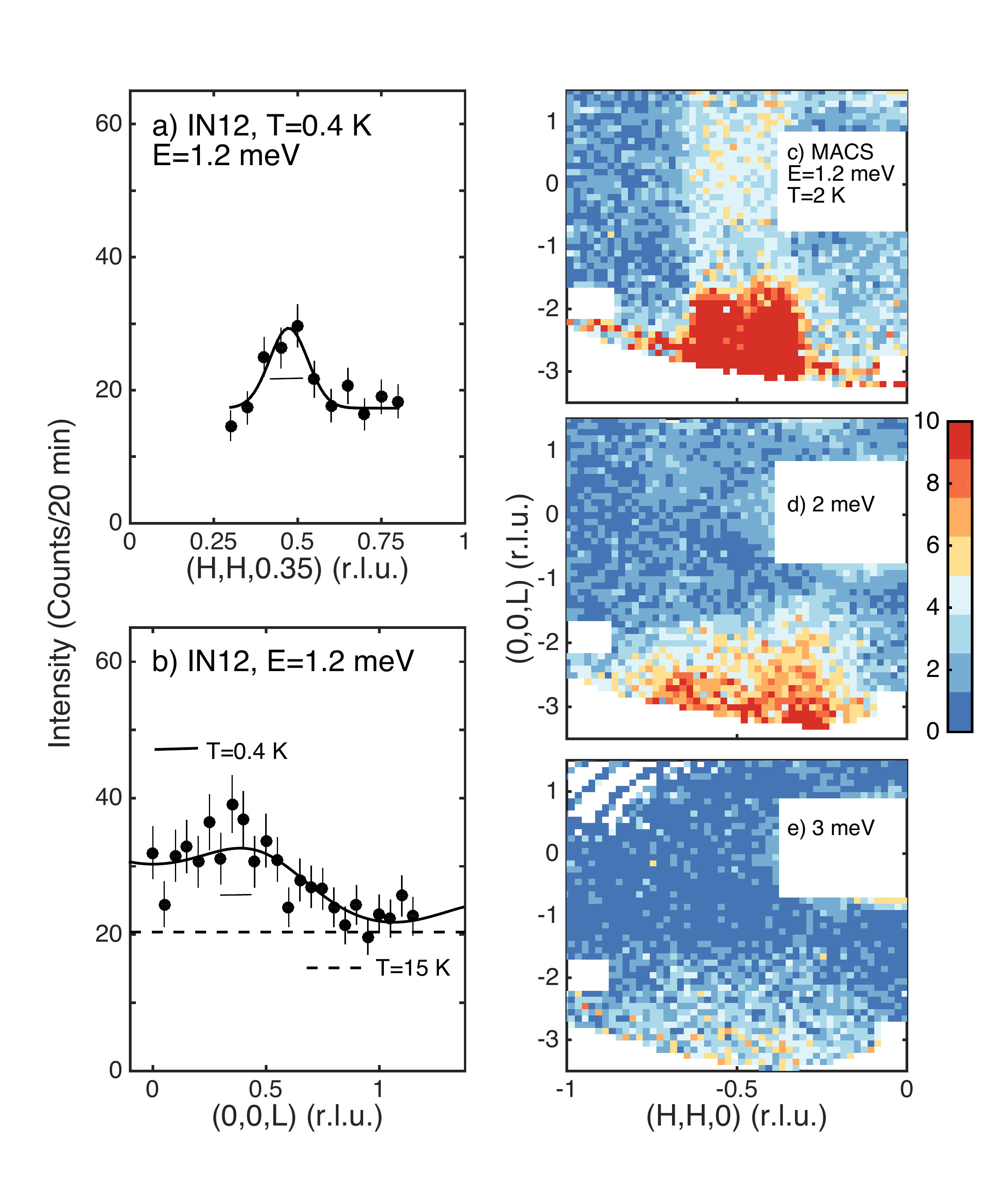}
\caption{\label{constantE} Constant energy scans taken on IN12 and MACS in the antiferromagnetic phase. (a)-(b) illustrate fluctuations polarized along $c$ with the horizontal bar being the spectrometer resolution.  (c)-(e) show constant energy slices showing the energy dependence of the spin fluctuations.  Fluctuations at large $L$ characteristic of predominately $a-b$ plane polarized fluctuations are present to high energy transfers.}
\end{figure}

We now discuss the inelastic scattering probing the dynamics.  Figure \ref{constantE} illustrates a summary of constant energy scans.   Figure \ref{constantE}  (a) shows a momentum scan along [110] finding the scattering to be peaked at (0.5,0.5) indicating antiferromagnetic correlations within the $a-b$ plane.  Figure \ref{constantE} (b) shows a scan along the [001] direction (corrected for absorption) finding momentum broadened correlations which decay rapidly with $L$.  The solid line is a fit to $I({\bf{Q}}) \propto f(Q)^{2} \times [1-({\bf{\hat{Q}\cdot \hat{c}}})^{2}] \sinh (c/\xi_{c})/[\cosh(c/\xi_{c})+\cos({\bf{Q \cdot c}})]$ which represents short-range antiferromagnetically correlated Ce$^{3+}$ moments polarized along $c$ with a dynamic correlation length $\xi_{c}$.  $f(Q)^{2}$ is the magnetic form factor.~\cite{Brown:tables} The dynamic correlation length was derived to be $\xi_{c}$= 3.1 $\pm$0.7 \AA\ indicating little coupling between the Ce$^{3+}$ layers.  The strong decrease in intensity with momentum transfer along $L$ illustrates that these fluctuations are predominately out of the $a-b$ plane (c-axis polarized) and hence referred to to as out-of plane fluctuations here (see supplementary information).  

\begin{figure}[t]
\includegraphics[width=8.0cm] {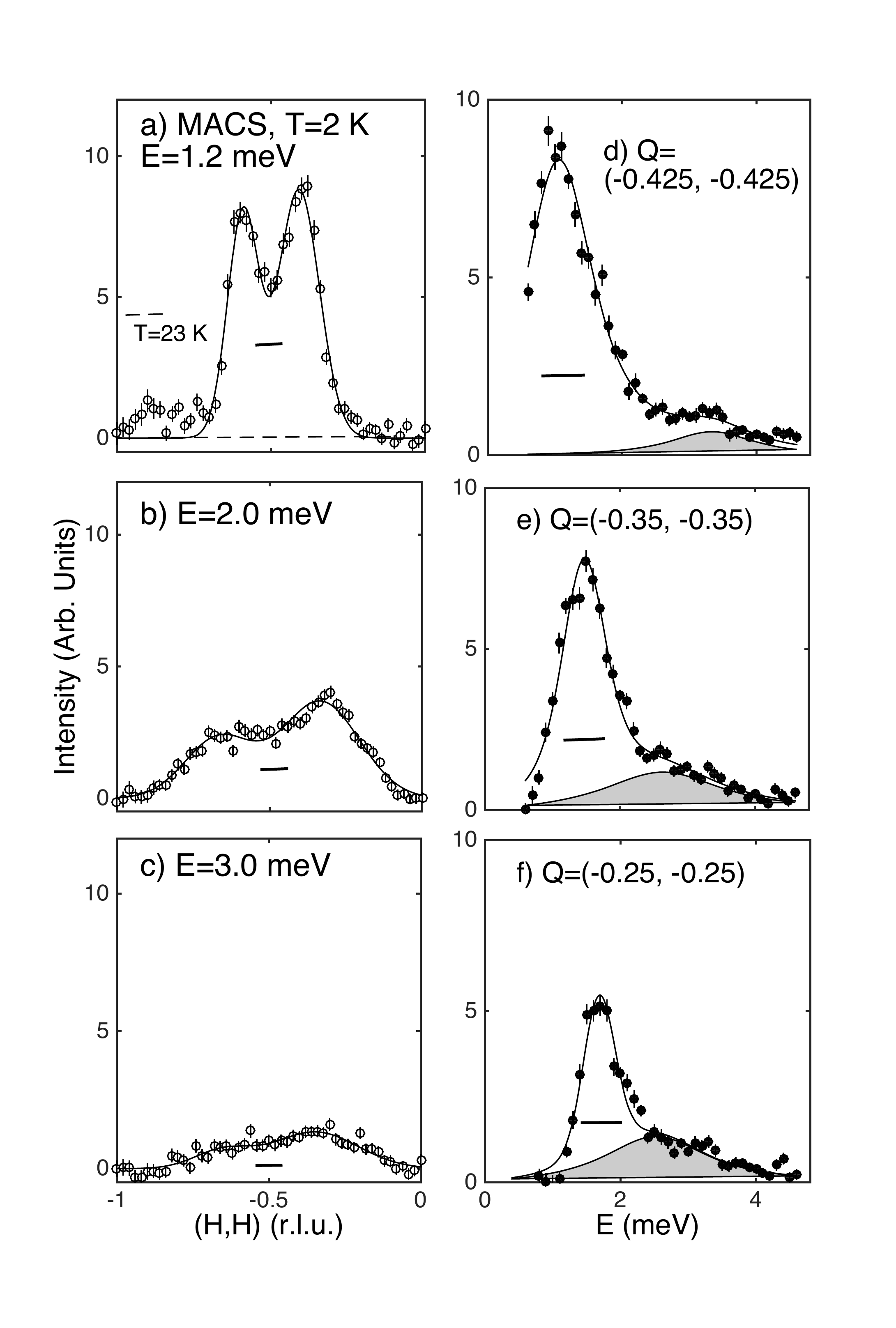}
\caption{\label{cuts} Constant energy (a-c) and $Q$ ($d-f$) cuts integrating over L=$[-4,-1.25]$ sensitive to predominately in-plane fluctuations.  The solid lines in (a)-(c) are to gaussians displaced from the commensurate (${1\over 2}$, ${1\over 2}$) position. Solid lines in (d)-(f) are fits to damped harmonic oscillators.   The shaded region is the broad heavily damped component.  (d-f) are integrated over $\pm(0.025,0.025)$.  The solid bars represent the experimental resolution.}
\end{figure}

Figures \ref{constantE} (c)-(e) illustrate full constant energy maps taken on MACS at energy transfers of 1.2-3 meV (c)-(e).  Panel (c) illustrates that, as well as the magnetic scattering near $L$=0 from the out of plane fluctuations, strong scattering is also present at large $L$ indicative fluctuations predominately polarized within the $a-b$ plane, referred to as in-plane fluctuations here.  We note that the energy transfer is significantly less than the first crystal field excitation at $\sim$ 7-9 meV, indicating that the transition results from excitations within the lowest energy Ce$^{3+}$ doublet.~\cite{Christianson02:66, Willers10:81}  The correlated scattering is present at higher energies as evidenced by similar scans in (d) and (e).  
  
\begin{figure}[t]
\includegraphics[width=8.0cm] {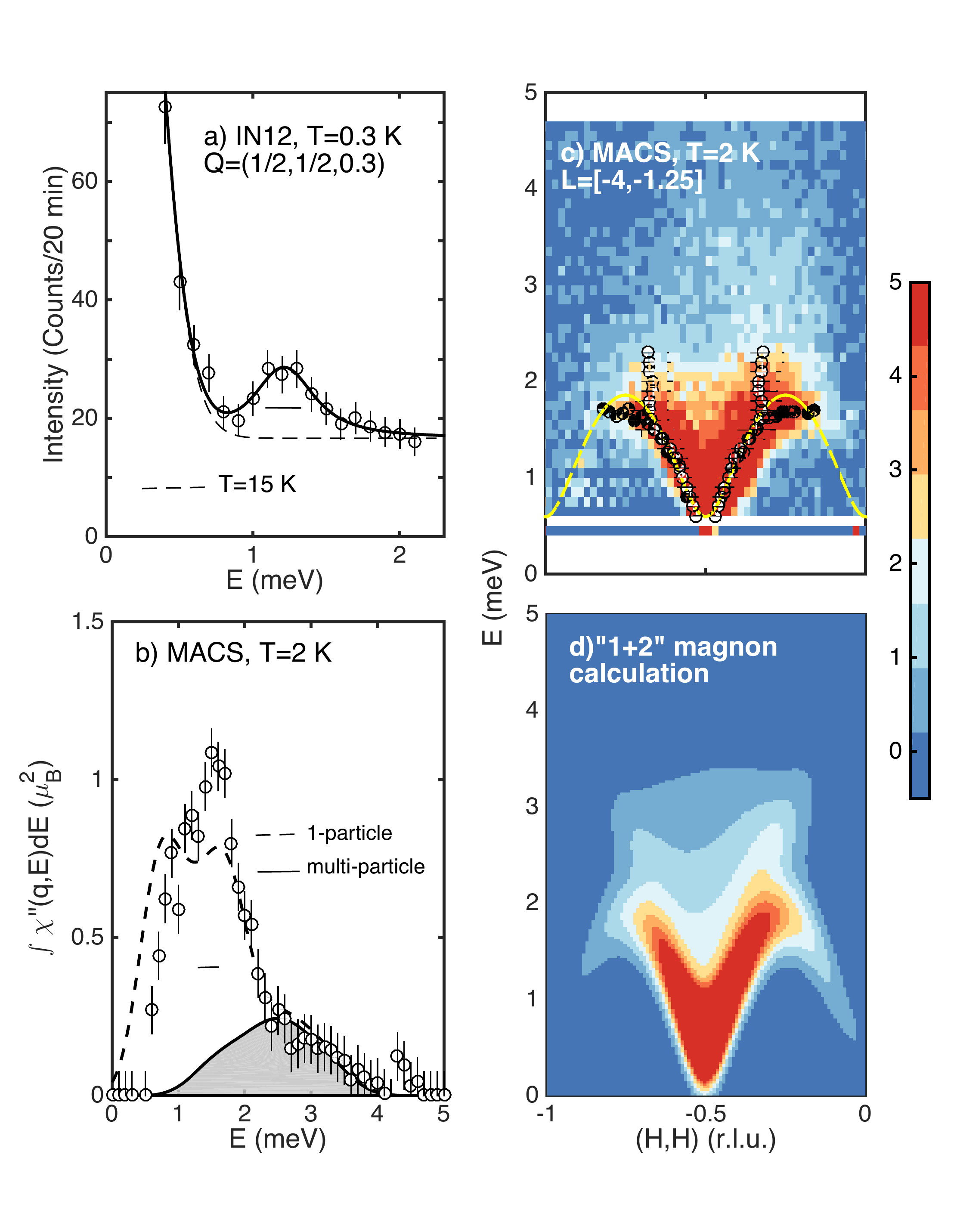}
\caption{\label{constantQ} Constant-$Q$ scans taken on IN12 and MACS. (a) illustrate the energy dependence of the $c$ axis polarized spin fluctuations.  (c) shows a constant-$Q$ slice taken on MACS (integrating $L$=[-4,-1.25]) with the solid points fits to constant-Q scans and the open circles fits to constant energy.  A continuum of scattering is present above the top of the ``1-magnon" band.  (d) shows a  calculation considering the parameterization in single (``1") and multiparticle (``2") states with the $\vec{Q}$ integrated intensities plotted in (b).}
\end{figure}

We now discuss the energy dependence.   Constant energy and momentum cuts are shown in Figs. \ref{cuts} (a)-(c) and (d)-(f) respectively.  As seen in both types of cuts, at low-energies the magnetic dynamics are described by two components - one which is sharp and resolution limited in energy and momentum and second higher energy component which is broadened in both momentum and energy.  

Figure \ref{constantQ} (c) displays a constant $Q$ slice (integrating over L=[-1.5,-4]) sensitive to the predominately in-plane scattering.   When all of the scattering is integrated over the magnetic Brillouin zone, the total spectral weight (accounting for absorption) is estimated at 2.0 $\pm$ 0.5 $\mu_{B}^{2}$ agreeing with expectations from single ion crystal field analysis (see supplementary information).     Both components need to to be considered to satisfy sum rules and obtain all of the required dynamic spectral weight.  

Neutron scattering is constrained by strict selection rules with the scattering process having $\Delta S_{z}=\pm$1 or 0.  Transverse spin excitations derive from harmonic theory and can be written as single quasiparticle or magnon excitations which are long lived in a magnetically ordered structure with resolution limited inelastic peaks.  Other anharmonic processes can occur including scattering from two magnons with opposite sign (ie. $\Delta S_{z}$=0 process) provided there is an interaction term between the single magnon quasiparticles in the Hamiltonian.  For collinear magnets, such terms are predicted to be weak from symmetry considerations, however for a noncollinear magnet, such as magnetic spiral or helix, such constraints are relaxed.~\cite{Zhitomirsky13:85,Villain74:35}  This additional cross section in the neutron response is constrained by momentum and energy conserving processes, and is possible over a wide range in energy and momentum which is determined by the single magnon dispersion.  Analogous classic examples of this cross section are found in model insulating low spin chains.~\cite{Tennant93:70,Tennant95:52,Lake09:06,Christiansen07:104,Stone06:440,Zal87:37,Kenzelmann01:87}  We now investigate whether the two component lineshape found here can be understood in terms of a single and multiparticle parameterization.  

We first consider the low-energy component of the cross section that is also resolution limited in energy.  Magnetic excitations for a planer helical magnet with a characteristic wavevector $\vec{q}_{c}$ are described by three modes with $\vec{Q}$=$\pm$$\vec{q}_{c}$ being in-plane modes and a commensurate mode describing out of plane fluctuations.~\cite{Chubukov84:17,Stock11:83,Coldea03:68,Dalidovich06:73}  

Fig. \ref{constantQ} (a) shows a constant $\vec{Q}$=(0.5,0.5,0.3) scan which is derived to have a strong $c$-axis polarized character.  An antisymmetric lorentzian fit gives a peak energy position of $\hbar\Omega$=1.21 $\pm$ 0.06 meV and line width (half-width) of $\hbar\Gamma$=0.22 $\pm$ 0.14 meV.   The out of plane fluctuations are therefore gapped as well as weakly dispersing.

To extract a dispersion and hence an estimate for the in-plane exchange interaction, we have fit constant energy scans (examples shown in Fig. \ref{cuts} (a)-(c)) to gaussians symmetrically displaced from the $\vec{Q}$=(${1 \over 2}$,${1 \over 2}$) and illustrated by the open circles in Fig. \ref{constantQ} (c).   The constant energy fits show dispersing excitations at wave vectors close to (${1 \over 2}$,${1 \over 2}$), but at the zone boundary near $({1\over 4},{1\over 4})$ the ``dispersion" becomes nearly vertical.   

Constant momentum scans in Fig. \ref{cuts} show that this vertical dispersion at the zone boundaries is due to the second short-lived and damped-in-energy component to the cross section.  To fully separate these two components, we have fit energy scans to two harmonic oscillators with one being resolution limited and the second damped in energy.  The sharp component is denoted by the filled circles in Fig. \ref{constantQ} (c).  To extract an estimate for the localized $J_{RKKY}$ exchange, we have fit the peak locations of the sharp component to the dispersion for a $j_{eff}={1 \over 2}$ ``spins" (capturing the doublet nature of the ground state) on a square lattice.  We have followed the classic model previously applied to Rb$_{2}$MnF$_{4}$ where a lattice periodic dispersion of $E(\vec{q})=2J_{RKKY}\sqrt{\alpha^{2}-\gamma(\vec{q})^2}$, with $\gamma(\vec{q})=\cos[\pi(H+K)]\cos[\pi(-K+H)]$ was used.  This provides a simple means of parametrizing the data and estimate of the nearest neighbor in-plane exchange.   We note that this model does not capture the out of plane mode which is found to show little dispersion and originate from weak coupling between the Ce$^{3+}$ layers.  Based on the fit in Fig. \ref{constantQ} to this parametrization, we extract $J_{RKKY}$=0.88 $\pm$ 0.05 meV and an anisotropy $\alpha$=1.06 $\pm$ 0.02 meV.  

Having described the sharp component sensitive to the antiferromagnetic exchange, we now discuss the broad continuum of scattering at higher energies.  We interpret and describe this component in terms of a multi magnon model termed the ``1+2" model.  The heavily damped features originates from unstable  particles where energy and momentum conservation result in a decay process.   As noted in Ref. \onlinecite{Chernyshev06:97}, the presence of the three modes imposed by the helical structure imply that excitations can decay into lower energy quasiparticles assuming there is a binding interaction.   For a given momentum transfer $\vec{k}$, the two particle excitations form a continuum of states and the energy and momentum positions where the cross section is finite is defined by conservation of momentum and energy.   Following the classical theory outlined in Ref. \onlinecite{Huberman05:72,Heilmann81:24} and using our parametrization of the single magnon scattering, we have calculated the energy and momentum dependence of the allowed multimagnon scattering. Fig. \ref{constantQ} (d) shows a plot of the scaled calculation with the one-magnon term superimposed to give the sharp component.  The momentum integrated intensity from the calculations is over plotted in Fig. \ref{constantQ} (b).  Deviations from calculations at low energies are likely due to experimental limitations owing to resolution, incoherent nuclear scattering, and absorption.   

Several features are reproduced in the multiparticle calculation: first, the broad continuum of scattering which extends up to nearly 2$\times J_{RKKY}$; and second, the nearly vertical columns of scattering which extend up in energy near the zone boundary.   Near the magnetic zone boundary, as illustrated in Fig. \ref{cuts}, the two components can be separated with both accounting for roughly equal amounts in terms of the integrated intensity.  The multiparticle model therefore provides an account of the neutron cross section once the single magnon component is parameterized with it giving the correct energy bandwidth and momentum dependence.  The multiparticle continuum is also predicted to have a longitudinal polarization,~\cite{Zhitomirsky13:85} consistent with the persistence to large $L$ shown in Fig. \ref{constantE}.

One thing that is not explicit in this analysis is how the coupling between single quasiparticles originates and what determines the relative spectral weight between the single and multiparticle components.  In insulating magnets, the spectral weight in the continuum comes from the Bragg peak, yet in CeRhIn$_{5}$, our analysis shows that the spectral weight draws from the inelastic component.  The symmetry of the helical magnetic structure simply implies that such multiparticle scattering is allowed in the neutron scattering cross section.  Such processes maybe determined by cubic terms in the Hamiltonian or possibly coupling resulting from the itinerant electronic nature of CeRhIn$_{5}$ as discussed elsewhere.~\cite{Macdonald88:37,Sandvik01:88,Hu90:41,Sylijuasen02:88}   However, we note that in classical and insulating magnets the multiparticle continuum is weak comprising $\sim$ 1-2 \% of the total spectral weight in Rb$_{2}$MnF$_{4}$.~\cite{Huberman05:72}  The relatively large size of the multiparticle continuum in CeRhIn$_{5}$ suggests that localized effects are not the cause and that the itinerant properties are important.   Our experiment suggests a low energy scale in CeRhIn$_{5}$ where the single quasiparticle description breaks down and interactions become important.    

The physics here might be more general and in particular, enhanced broadening in the neutron cross section has been observed near the zone boundary in metallic Fe$_{1+x}$Te~\cite{Stock14:xx}, and the cuprates YBa$_{2}$Cu$_{3}$O$_{6.35}$~\cite{Stock07:75,Stock10:82}, La$_{2}$CuO$_{4}$~\cite{Headings10:105}, and Sr$_{2}$CuO$_{2}$Cl$_{2}$.~\cite{Plumb14:89}  These might indicate an interaction similar that discussed here yet much weaker due to symmetry constraints determined by the collinear structures.  An alternate view is that the continuum in CeRhIn$_{5}$ results from the single magnon branch at low energies interacting with a continuum of electronic excitations as suggested in itinerant ferromagnets magnets such as MnSi~\cite{Ishikawa85:31} and Fe~\cite{Paul88:38}.  However, this scenario results in the disappearance or strong dampening of the single magnon branch and not the presence of two distinct components observed here in CeRhIn$_{5}$.  This high energy continuum may represent a direct measure of the hybridization gap which characterizes the energy scale where the quasiparticles cross over from localized to itinerant and such energy scales are expected to be on the order of $\sim$ meV in CeRhIn$_{5}$.~\cite{Park08:105}

In summary, we have studied the excitations in helical CeRhIn$_{5}$ and found the presence of a strong continuum along with sharp single magnon excitations.   Given both components are required to satisfy neutron scattering sum rules, we understand the cross section in terms of a ``1+2" particle model where the broad component originates from multiparticle states with the energy and momentum dependence fixed by energy and momentum conservation laws determined by the single magnon cross section.  We propose the multiparticle component originates from coupled magnonss, observable given the relaxed symmetry constraints from a helical magnet.   Our measurements directly observe the breakdown of a the single quasiparticle, or magnon, picture for an itinerant magnet.

This work was funded by the Carnegie Trust for the Universities of Scotland, the Royal Society, the Royal Society of Edinburgh, the STFC, and the EPSRC (M01052X).  Part of this work was carried out at the Brookhaven National Laboratory which is operated for the U. S. Department of Energy by Brookhaven Science Associates (DE-AcO2-98CH10886).

%\bibliography{CeRhIn5_excitations}

%merlin.mbs apsrev4-1.bst 2010-07-25 4.21a (PWD, AO, DPC) hacked
%Control: key (0)
%Control: author (8) initials jnrlst
%Control: editor formatted (1) identically to author
%Control: production of article title (-1) disabled
%Control: page (0) single
%Control: year (1) truncated
%Control: production of eprint (0) enabled
%

\end{document}

\title{Supplementary information for ``Single to multi quasiparticle excitations in the itinerant helical magnet CeRhIn$_{5}$"}

\author{C. Stock}
\affiliation{School of Physics and Astronomy, University of Edinburgh, Edinburgh EH9 3JZ, UK}
\author{J. A. Rodriguez-Rivera}
\affiliation{NIST Center for Neutron Research, National Institute of Standards and Technology, 100 Bureau Dr., Gaithersburg, MD 20899}
\affiliation{Department of Materials Science, University of Maryland, College Park, MD  20742}
\author{K. Schmalzl}
\affiliation{Julich Centre for Neutron Science, Forschungszentrum Julich, Outstation at Institut Laue-Langevin, Boite Postale 156, 38042 Grenoble Cedex 9, France}
\author{E. E. Rodriguez}
\affiliation{Department of Chemistry of Biochemistry, University of Maryland, College Park, MD, 20742, U.S.A.}
\author{A, Stunault}
\affiliation{Institute Laue-Langevin, B. P. 156, 6 rue Jules Horwitz, F-38042 Grenoble Cedex 9, France}
\author{C. Petrovic}
\affiliation{Department of Physics, Brookhaven National Laboratory, Upton, New York, 11973, USA}

\date{\today}

\begin{abstract}

Supplementary information is provided in support of the main text discussing the magnetic structure and excitations in CeRhIn$_{5}$.  Further information is given regarding (1) experimental details of the neutron scattering measurements; (2) symmetry analysis of the magnetic structure;  (3) temperature dependence of the inelastic scattering; (4) model lineshapes used to parametrize the dynamic response; (5) crystal fields and matrix elements; (6) single and multiparticle model imposing energy and momentum conservation; and (7) details on absorption estimates using a gaussian integration analysis.

\end{abstract}

\maketitle

\section{Further Experimental details}

To formulate a complete picture of the magnetic structure and dynamics, several different elastic and inelastic experiments were performed.  The experimental details are provided here.

\subsection{Inelastic scattering}

Neutron inelastic scattering measurements were performed on the IN12 cold triple-axis spectrometer (ILL, Grenoble, France) and using the MACS cold triple-axis (NIST, Gaithersburg, USA).  On IN12, probing the $c$-axis polarized spin fluctuations, the final neutron energy was fixed at E$_{f}$=4.6 meV and a Be filter was used on the scattered side to filter out higher order harmonics.   In-plane fluctuations were studied using the MACS spectrometer on the BT9 beam port at NIST with E$_{f}$=5.05 meV and with cold Be filters on the scattered side.  Details on MACS are provided in Ref. \onlinecite{Rodriguez08:19}.  For all inelastic measurements,  a 6.1 g plate was cut from a larger boule to minimize neutron absorption effects while retaining structural integrity for mounting.  Further analysis of the absorption effects are provided at the end of this supplementary information.  

\subsection{Elastic scattering}

Unpolarized diffraction studies were performed on the D23 diffractometer (incident $\lambda$=2.385 \AA) and polarized work was done on the D3 diffractometer (incident $\lambda$=0.825 \AA) with cryopad installed.  As noted in the main text, this work largely confirmed the published magnetic structure.  For all elastic measurements, plate-like samples with masses ranging from 25-75 mg where used.    Further details on the cryopad technique and the experimental setup and be found in Ref. \onlinecite{Tasset99:267}.

\section{Further details regarding neutron diffraction}

In this section, we provide additional information regarding the magnetic symmetry and experiments we performed on CeRhIn$_{5}$ probing the critical properties near T$_{N}$.

\subsection {Magnetic symmetry}

To provide a starting point for describing the magnetic structure and fitting the polarized neutron results, we discuss the magnetic symmetry of CeRhIn$_{5}$.  The irreducible representations and basis vectors were calculated using the the programmes SARAh and Mody and shown in Table \ref{irrep}.  There are two irreducible representations for the space group $P 4/m m m$ with magnetic ordering wavevector ${\bf{q}_{0}}$=(0.5,0.5,$\alpha$); $\tau_{2}$ which has two basis vectors ($\psi_{2,3}$) along $[100]$ and $[010]$ and $\tau_{1}$ with $\psi_{1}$ along  [001].   One irreducible representation ($\tau_{2}$) contains two basis vectors ($\psi_{2,3}$) within the $a-b$ plane while the second ($\tau_{1}$) has a vector ($\psi_{1}$) along $c$.  

\begin{table}
\caption{The basis vectors of the irreducible representations (I$_{rrep}$) for the space group $P 4/m \ 2/m \ 2/m$ (No. 123) with magnetic ordering wavevector ${\bf{q_{0}}}=( {1 \over 2}, {1 \over 2} ,\delta)$.}
\label{irrep}
\begin{tabular}{clclclclcl}
\hline
\hline
Irrep 		& &    Basis Vector  	& & 	m$_{x}$ 		& & 		m$_{y}$ 		& &  m$_{z}$ \\
\hline
$\tau_1$  & & 	$\psi_{1}$ 		& & 	0 					& &  		0 					& & 1                                         \\
\hline
$\tau_2$  & & 	$\psi_{2}$ 		& &  1  					& & 		0  				& & 0               \\
 				& & 	$\psi_{3}$ 		& &  	0  				& & 		-1 				& & 0               \\
\hline
\hline
\end{tabular}
\end{table}

The fit provided in Fig. 1 in the main text assumes a perfect $a-b$ spiral with ${\bf{M}}={\bf{M}}_{a}+i{\bf{M}}_{b}$ (with $|{\bf{M_{a}}}|=|{\bf{M_{b}}}|$).  This structure is based on equal weights of the $\tau_{2}$ basis vectors.  Such a magnetic structure gives a polarization matrix with off diagonal elements $P_{21,31}=\eta{{M_{y}M_{z}}\over{M^{2}}}$, where $\eta$ is the domain population difference given by ($\eta \equiv {{V_{+}-V_{-}} \over {{V_{+}+V_{-}}}}$, with $V_{\pm}$ the volume fractions of the two $\pm$ helical domains), and diagonal elements $P_{22,33}={{M_{y}^{2}-M_{z}^{2}}\over{M^{2}}}$.  

The polarized neutron results outlined in the text are consistent with previous unpolarized work and observe a finite P$_{21,31}$ matrix indicating a helical magnetic structure with a net domain imbalance.  Symmetry consideration would imply two domains with a vector chirality pointing along $\pm$ [001] and Fig. 1 (b) is fitted with a domain population of $\eta$=0.68 $\pm$ 0.05, likely the result of strain and sample geometry.   This domain preference  allows us to uniquely identify the structure as a spiral over a spin density wave.  Fig. 1$b)$ is a fit to all nonzero matrix elements using only the $\tau_{2}$ irreducible representation with equal components of the two basis vectors (magnetic structure illustrated in Fig. 1 (a).  No detectable $c$-axis component of the moment was found indicating the absence of basis vectors from $\tau_{1}$.  

The in plane spiral derived from spherical polarimetry is consistent with only one irreducible representation ($\tau_{2}$) being involved in the transition.  Landau theory then predicts a second order phase transition consistent with the experimental results (Fig. \ref{order_param}) for the order parameter $m^{2}$ which is proportional to the neutron scattering cross section.

\renewcommand{\thefigure}{S1}
\begin{figure}[t]
\includegraphics[width=8.6cm] {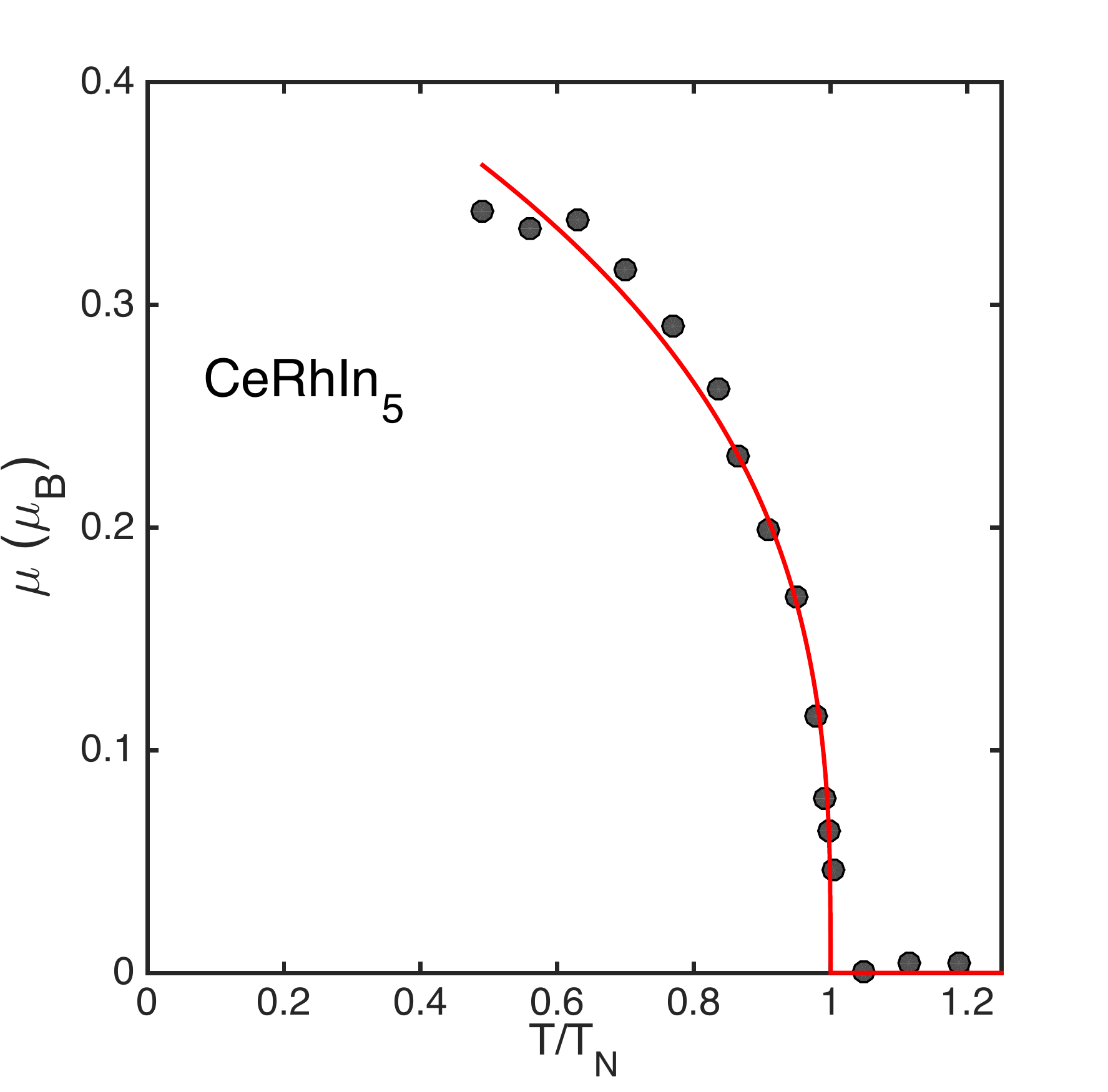}
\caption{\label{order_param} The magnetic order parameter for CeRhIn$_{5}$ illustrating the second order nature of the phase transition.  This supports the fitted magnetic structure as discussed using symmetry analysis in this supplementary information.}
\end{figure}

\section{Further details on neutron inelastic scattering experiment}

In this section, we discuss additional information and experiments on the magnetic dynamics.  We first describe how we verified the magnetic origin through the temperature dependence and then the line shapes and models we have used to parametrize the neutron inelastic scattering data.

\subsection{Temperature dependence as confirmation of magnetic origin}

\renewcommand{\thefigure}{S2}
\begin{figure}[t]
\includegraphics[width=8.5cm] {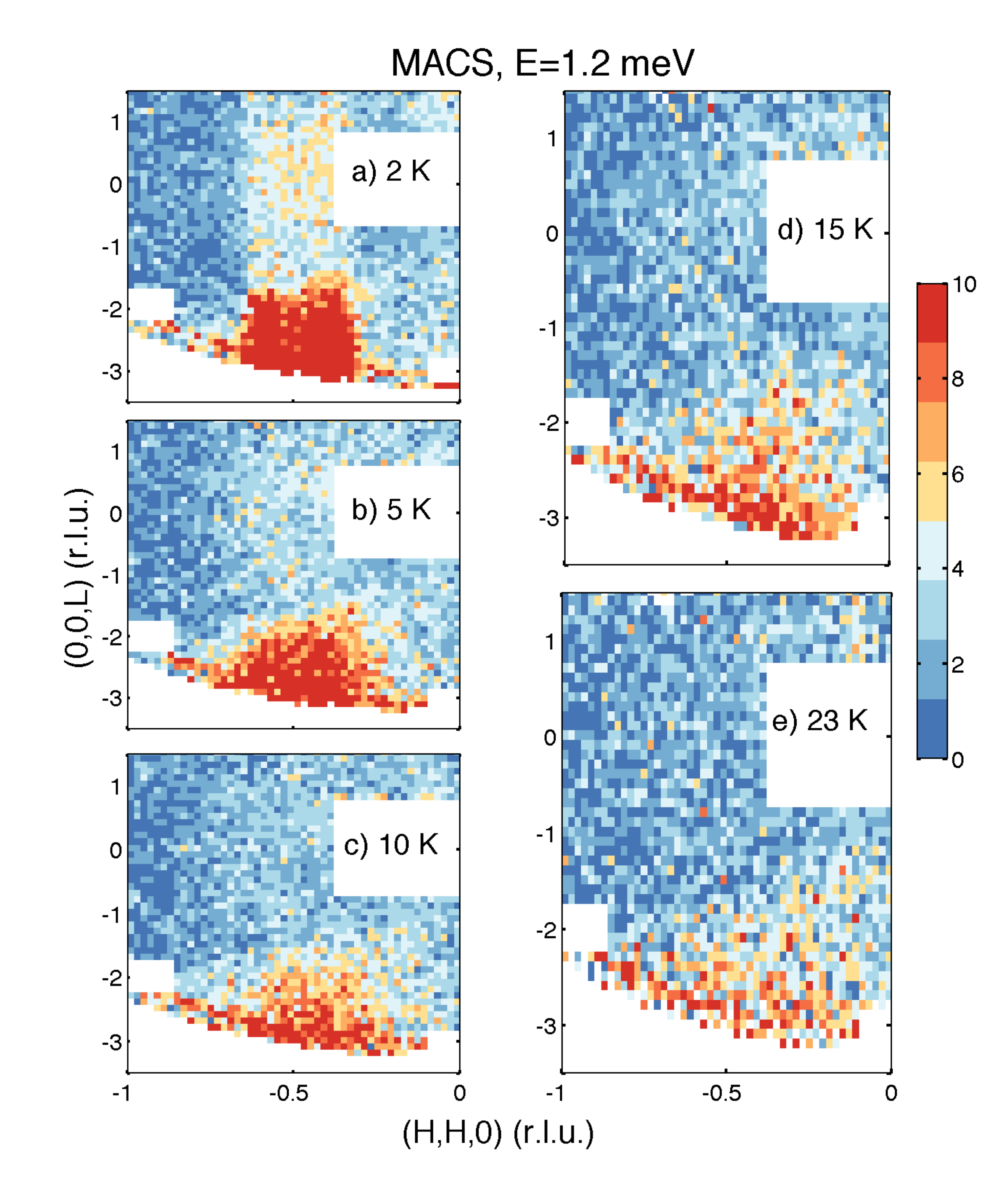}
\caption{\label{temp} The temperature dependence of the magnetic scattering probed through constant energy slices at E=1.2 meV.  The white squares are regions in momentum space masked due to incomplete subtraction from Bragg tails leaking into the inelastic channel.  The masking was necessary for computing the integrated intensity.}
\end{figure}

Magnetic scattering typically decays with increasing momentum transfer, however a competing factor is the orientation factor in the neutron scattering cross section which imposes the fact that neutrons are sensitive to the magnetic moment perpendicular to the momentum transfer.  To verify that the scattering we observe in the main text of the paper is magnetic we have measured the temperature dependence.  Figure \ref{temp} illustrates the temperature dependence measured on MACS of the in plane fluctuations.  Constant energy slices at E=1.2 meV illustrate that the magnetic scattering decays rapidly with temperature and is nearly totally suppressed at 23 K.  This provides additional confirmation that the scattering studied is indeed magnetic in origin.

\subsection{Lineshape for magnetic excitations}

To fit and separate the sharp spin fluctuations from the broad continuum scattering in the main text, we have fit the data to a linear combination of damped harmonic oscillators.  Concerning the notation we note that the measured neutron scattering intensity is directly proportional to $S({\bf{Q}},E)$, 

\begin{eqnarray}
I({\bf{Q}},E) \propto S({\bf{Q}},E) \equiv {1\over \pi} [n(E)+1] \chi''({\bf{Q}},E).
\label{intensity}
\end{eqnarray}

\noindent We have set the imaginary part of the susceptibility to be a linear combination of lorentzians.  This particular form of the scattering function follows detailed balance.

\begin{eqnarray}
\chi''(E)= \chi'_{0} \left( {1\over {1+\left( { {E-\hbar \Omega_{0}} \over \Gamma} \right)^{2}}}- {1\over {1+\left( { {E+\hbar \Omega_{0}} \over \Gamma} \right)^{2}}} \right).
\label{harmonic}
\end{eqnarray}

\noindent  This particular line shape was chosen to describe both the sharp (``1-magnon") and broad multiparticle (``2-magnon") continuum owing to the quality of the fits in Fig. 3 of the main text and also the ability to integrate the line shape to compare with the expected matrix elements from crystal field theory discussed below.  We note that the ``1-magnon" component was found to be resolution limited and the multiparticle component was found to be significantly damped.

\subsection{Crystal Fields and integrated intensities}

In our analysis and interpretation of the data we have discussed the excitations in terms of transitions within the ground state doublet.  To make this assumption and to support this discussion, an understanding of the crystal fields is required and we provide an analysis of this in this section.

In the main text of the paper we discuss the absolute intensity for both the static and dynamic components.  For a theoretical estimate of the integrated intensities which depend on transition matrix elements, we have used the crystal field parameters provided in Ref. \onlinecite{Christianson02:66,Willers10:81} for CeRhIn$_{5}$.  We note that both Ref. \onlinecite{Christianson02:66} and Ref. \onlinecite{Willers10:81} give consistent answers for the ground state wavefunctions in CeRhIn$_{5}$ and therefore we have used the average value in our analysis.  Based upon symmetry considerations, the Hamiltonian for Ce$^{3+}$, J=$5/2$ in a tetragonal crystal field is given by the following equation in terms of Steven's operators ($O_{2}^{0}$, $O_{4}^{0}$, and $O_{4}^{4}$) and parameters ($B_{2}^{0}$, $B_{4}^{0}$, and $B_{4}^{4}$).~\cite{Hutchings65:45}

\begin{eqnarray}
H_{CEF}=B_{2}^{0}O_{2}^{0}+B_{4}^{0}O_{4}^{0}+B_{4}^{4}O_{4}^{4}
\label{ham}
\end{eqnarray}

\begin{table}[h!]
\caption{Stevens coefficients taken from Ref. \onlinecite{Christianson02:66} and \onlinecite{Willers10:81}.  Averages are given in the last column. }
\begin{tabular}{c|c|clcl}
 \hline
$B_{2}^{0}$ &-1.03 meV & -0.928 meV & $\equiv$ -0.98 meV \\
$B_{4}^{0}$ & 0.044 meV & 0.052 meV & $\equiv$ 0.048 meV \\
$B_{4}^{4}$ & 0.122 meV & 0.128 meV & $\equiv$ 0.125 meV \\
 \hline
\end{tabular}
\label{table_cef}
\end{table}

The eigenvalue spectrum of this Hamiltonian illustrates that the ground state is a doublet (expected from Kramer's theorem) and the excitation spectrum consists of a further two doublets at finite energy transfer.  The parameters have been analyzed experimentally in Ref. \onlinecite{Christianson02:66} and are listed in Table \ref{table_cef}.  Substituting these back into Eqn. \ref{ham} and solving for the eigenevectors and eigenvalues allows us to calculate the cross sections for transitions between levels.  Of particular interest, in this experiment, is the intensity for transitions within the ground state doublet (giving the inelastic cross section) and also the ordered moment (elastic neutron cross section).  We calculate the following matrix elements.

\begin{eqnarray}
 \sum_{i=x,y} |\langle -|J_{i}|+\rangle|^{2} \mu_{B}^2=2.0 \mu_{B}^2 \nonumber \\
g_{J}^2|\langle 0|J_{z}| 0\rangle|^{2} \mu_{B}^{2}= 0.38 \mu_{B}^2.\nonumber
\end{eqnarray}

\noindent The matrix elements $|\langle -|J_{i}|+\rangle|$ correspond to inter doublet transitions and $|\langle 0|J_{z}|0\rangle|$ represents transitions within a given member of the ground state doublet.  The matrix element for $J_{z}$ is in reasonable agreement with the ordered moment cited in Ref. \onlinecite{Bao00:62} and found in the main text in Fig. 1 ($\sim$ 0.35 $\mu_{B}^{2}$).  The total fluctuating integrated moment discussed in the main text is 2.0 $\pm$ 0.5 $\mu_{B}^{2}$ and is close to expectations based on the single-ion crystal field analysis.  We emphasize that the large errorbar here is due to non trivial corrections from absorption.  Nevertheless, the analysis illustrates that most of the spectral weight is captured by the combination of single and multi particle scattering discussed above.

\subsection{Separation of in and out of plane fluctuations}

\renewcommand{\thefigure}{S3}
\begin{figure}[t]
\includegraphics[width=8.5cm] {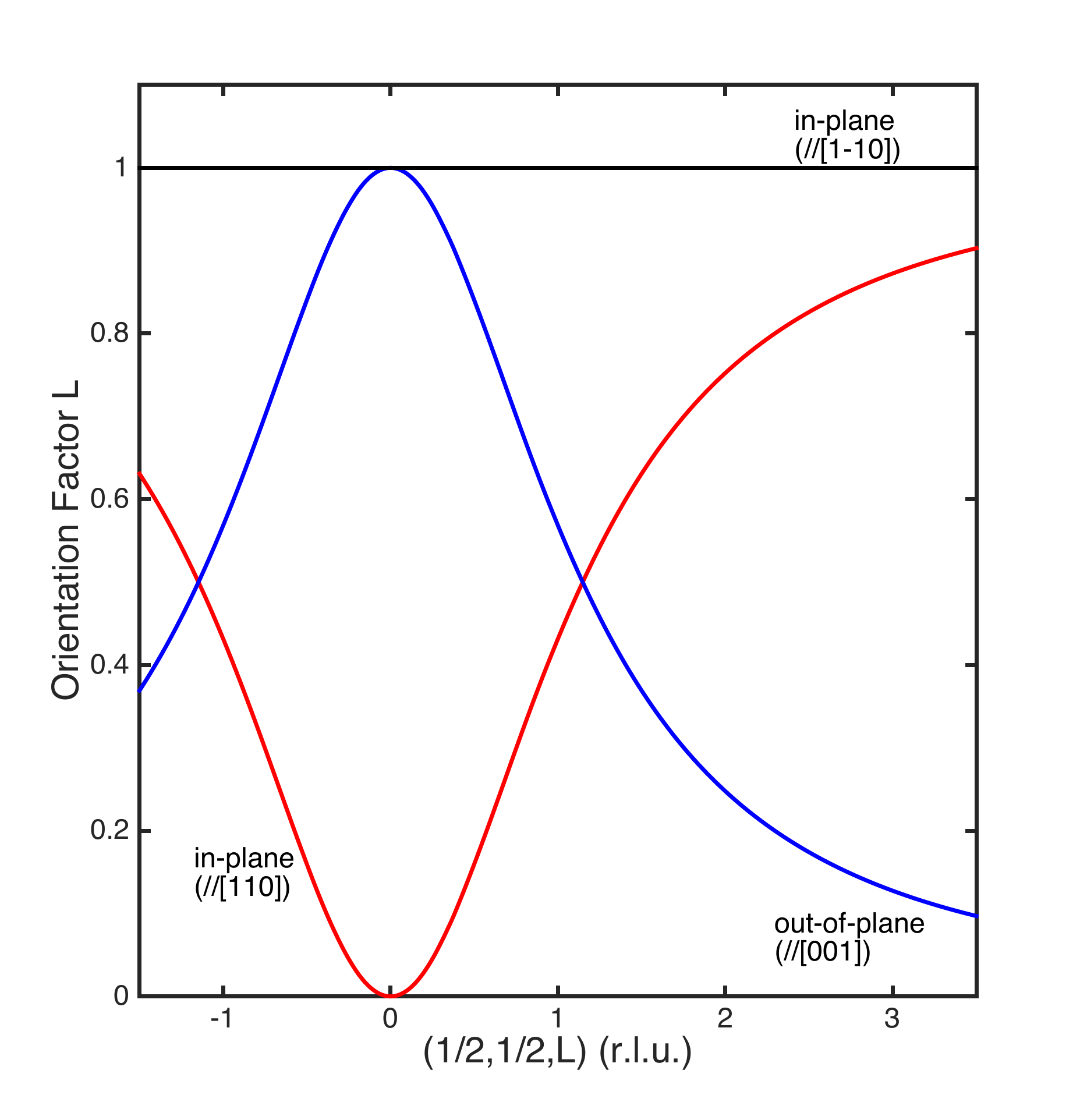}
\caption{\label{orient}  The orientation factor for magnetization in the plane and out of the plane plotted as a function of L at (0.5,0.5,L). Based on the $L$ dependence, we can derive whether the fluctuations are fluctuation in our out of the $a-b$ plane.}
\end{figure}
One of the key components of the analysis presented in the main text is the discussion of fluctuations polarized within the $ab$ plane (denoted as in-plane) and along $c$ (denoted as out of plane).  While polarized inelastic neutron scattering is the ideal technique for resolving the polarization direction for magnetic fluctuations, the use of polarized neutrons in absorbing samples and also for studying weak continuum scattering is intensity limited and not always possible.  By studying the momentum dependence of the fluctuations, the polarization can be determined through the orientation factor in neutron scattering which states that neutrons are sensitive to the moment perpendicular to to the momentum transfer, 

\begin{eqnarray}
L(\vec{Q})=1-(\hat{Q}\cdot\hat{M})^{2}.
\label{orient}
\end{eqnarray}

\noindent Where the unit vector $\hat{M}$ is the polarization direction and  $\hat{Q}$ is the direction of momentum transfer.  Scans along L are particularly sensitive to the polarization direction as described in the text.  Figure \ref{orient} plots the orientation factor along the (0.5, 0.5, $L$) direction for in and out of plane polarizations.   In particular, scattering decaying rapidly with $L$ is indicative of out-of-plane fluctuations.  Based on the strong decay of intensity for out-of-plane fluctuations, we associate the scattering at large $L$ discussed in the main text originating from in-plane fluctuations. 

\subsection{``1+2" magnon calculations}

The magnetic dynamics we observe consist of two distinct components - a temporally sharp portion which corresponds to harmonic propagating excitations and a second continuum component which is broad in momentum and energy.  To parameterize the spin dynamics, we have followed the classical case of insulating Rb$_{2}$MnF$_{4}$ which consists of classical large S=${5 \over 2}$ spins coupled on a two dimensional square lattice.  Below we outline the formula and notation used in the main text to parametrize the data.  The discussion largely follows that in Ref. \onlinecite{Huberman05:72}.

\subsubsection{One magnon scattering}

The sharp component which gives a lower bound to the magnetic scattering in energy transfer has a clear dispersion in energy and momentum.  The line shape is also nearly resolution limited based on the fits to a harmonic oscillator described above.   To extract a coupling between the Ce$^{3+}$ sites we have fit the dispersion to a ``one-magnon" cross section corresponding to transverse spin fluctuations given by,

\begin{eqnarray}
S^{xx,yy}({\bf{Q}},E)\propto {1\over 2} (u({\bf{Q}})+v({\bf{Q}}))^{2}\delta(E-E({\bf{Q}})).
\label{2magnon}
\end{eqnarray}

\noindent Here $E({\bf{Q}})$ is the dispersion, $E({\bf{Q}})=2J_{RKKY} \sqrt{\alpha^{2}-\gamma({\bf{Q}})}$ and where $u({\bf{Q}})=\cosh \theta$,  $v({\bf{Q}})=\sinh \theta$, and $\tanh 2\theta=-\gamma({\bf{Q}})/\alpha$.  In this analysis we have taken the fact that the magnetic excitations are at energies well below the first excited doublet (from a crystal field analysis) and therefore we observe excitations within the ground state doublet.  We have therefore taken $S_{eff}={1\over 2}$ in the above analysis to account for the doublet nature of the ground state.

\subsubsection{Two magnon scattering}

Having discussed how we accounted for the sharp component, we now discuss the broad continuum excitation which we have interpreted in terms of multiparticle states and as a guide we have employed a ``2-magnon" cross section to understand it.  The longitudinal fluctuations which reduce the spectral weight present in the elastic and sharp 1-magnon contribution described above can be described in terms of 2-magnon scattering.  As described in Refs. \onlinecite{Huberman05:72,Heilmann81:24}, this can be written by the following equation.

\begin{eqnarray}
S^{zz}({\bf{Q}},E)=\nonumber...\\
{1\over {2N}} \sum_{\bf{\tilde{Q}}} \left( u({\bf{\tilde{Q}}})v({\bf{Q}}-{\bf{\tilde{Q}}})-v({\bf{\tilde{Q}}})u({\bf{Q}}-\bf{{\tilde{Q}}}) \right) \nonumber...\\
\delta\left(E-E({\bf{Q}})-E({\bf{\tilde{Q}}}-{\bf{Q}}) \right),
\label{2magnon}
\end{eqnarray}

\noindent where $u({\bf{Q}})$ and $v({\bf{Q}})$ are given above.  To account for the fact that the fluctuations are polarized along the magnetic moment direction, we have included an orientation factor of in-plane fluctuations.  In Fig. 3 of the main text we have present a model calculation of both one and two magnon components individually scaled to mimic experiment.  The observed inelastic spectrum in CeRhIn$_{5}$ is therefore well described by an anomalously large multiparticle component.  Possible origins of this were discussed in the main text.

\section{Absorption correction}

\renewcommand{\thefigure}{S4}
\begin{figure}[t]
\includegraphics[width=8.5cm] {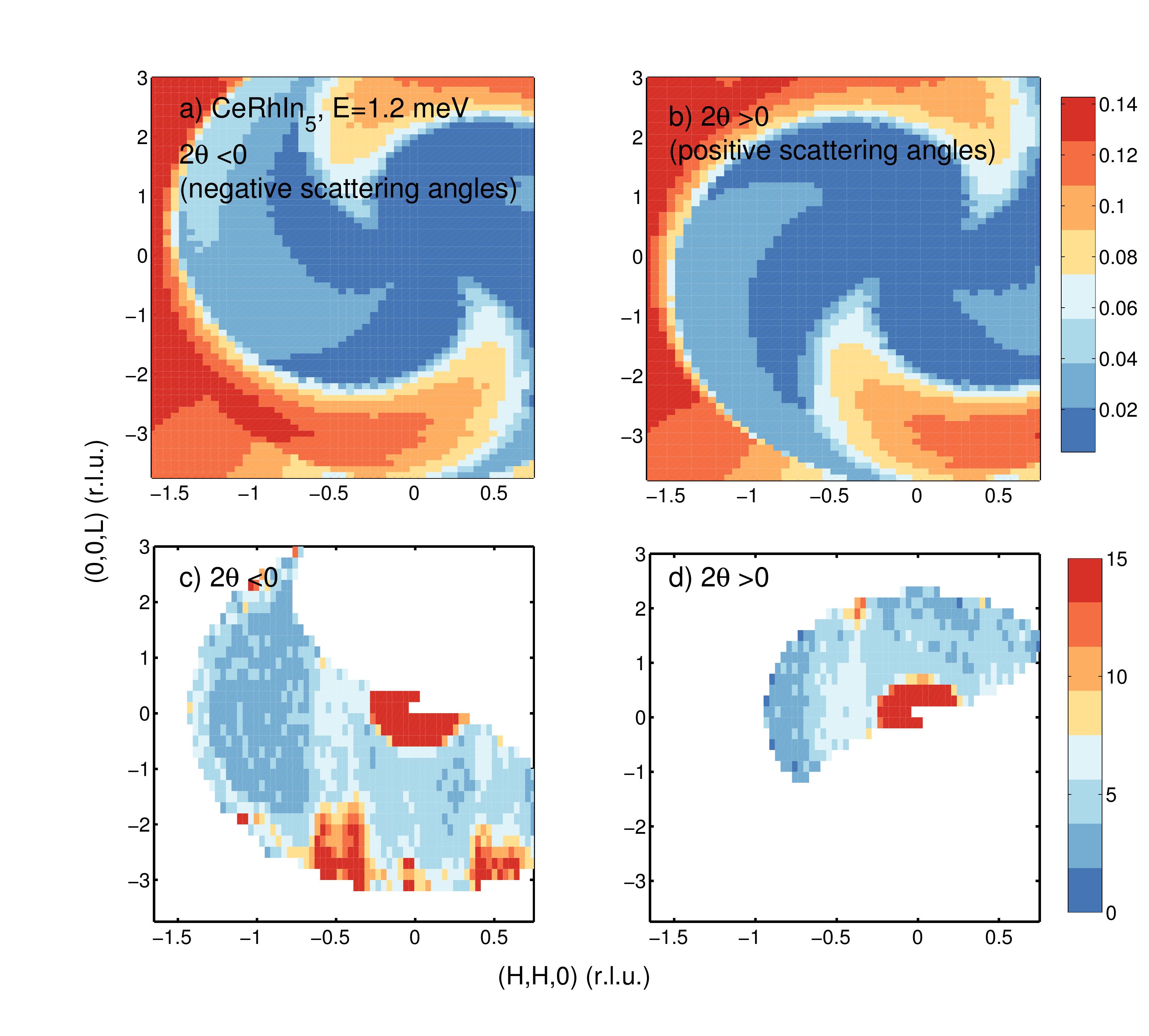}
\caption{\label{absorb} $a-b)$ Calculated transmission at E=1.2 meV for scattering angles ($2\theta$) both positive and negative. Sample cuts using negative ($c$) and positive ($d$) scattering angles.  The experimental data probing the dispersion in the main text used negative scattering angles ($2\theta<$0).}
\end{figure}

Both rhodium (Rh) and indium (In) are large neutron absorbers with absorption cross sections of 144.8(7) and 193.8 (1.5) barns respectively.~\cite{Sears92:3}  The cross section is particularly large in comparison to the magnetic cross section dictated by $(\gamma r_{0})^2/4=73$ mbarns. The sample used for the experiment was a 6.5 mm thick plate of CeRhIn$_{5}$ which gave a strong momentum dependence to the transmission.  Given the strong angular dependence of the absorption correction, to guide the experiment we performed a gaussian integration analysis based on the algorithm in Ref. \onlinecite{Wuensch65:122} where the transmission is defined as $T={1\over M} \sum_{m=1}^{M} e^{-\mu(t_{0}+t)}$.  A slice of the calculated transmission coefficient at E=1.2 meV is shown in Fig. \ref{absorb} for both positive and negative scattering angles (Fig. \ref{absorb} (a)-(b).  The experimental results (uncorrected for background) are given in panels (c)-(d) for negative and positive scattering angles respectively.  Given the angular dependence of the transmission, we focussed our experiment for negative scattering angles and at large values of $L$.  More details of this finite element analysis applied to absorbing systems will be presented elsewhere.

%\bibliography{CeRhIn5_excitations}

%merlin.mbs apsrev4-1.bst 2010-07-25 4.21a (PWD, AO, DPC) hacked
%Control: key (0)
%Control: author (8) initials jnrlst
%Control: editor formatted (1) identically to author
%Control: production of article title (-1) disabled
%Control: page (0) single
%Control: year (1) truncated
%Control: production of eprint (0) enabled
%